\documentclass[a4paper]{jpconf}
\usepackage{graphicx}
\begin{document}

\title{Addendum to `Mapping electron dynamics in highly transient EUV photon-induced plasmas: a novel diagnostic approach using multi-mode microwave cavity resonance spectroscopy'}

\author{B.\ Platier}
\address{Department of Applied Physics, Eindhoven University of Technology, PO Box 513, 5600MB Eindhoven, The Netherlands}
\author{F.M.J.H.\ van de Wetering}
\address{ASML, The Netherlands, PO Box 324, 5500AH Veldhoven, The Netherlands}
\author{M.A.W.\ van Ninhuijs}
\address{Department of Applied Physics, Eindhoven University of Technology, PO Box 513, 5600MB Eindhoven, The Netherlands}
\author{G.J.H.\ Brussaard}
\address{ASML, The Netherlands, PO Box 324, 5500AH Veldhoven, The Netherlands}
\author{V.Y.\ Banine}
\address{Department of Applied Physics, Eindhoven University of Technology, PO Box 513, 5600MB Eindhoven, The Netherlands}
\address{ASML, The Netherlands, PO Box 324, 5500AH Veldhoven, The Netherlands}
\author{O.J.\ Luiten}
\address{Department of Applied Physics, Eindhoven University of Technology, PO Box 513, 5600MB Eindhoven, The Netherlands}
\author{J.\ Beckers}
\address{Department of Applied Physics, Eindhoven University of Technology, PO Box 513, 5600MB Eindhoven, The Netherlands}
\ead{b.platier@tue.nl}

\begin{abstract}
	\noindent A new approach for an in-line beam monitor for ionizing radiation was introduced in a recent publication (Beckers, J., et al.\ "Mapping electron dynamics in highly transient EUV photon-induced plasmas: a novel diagnostic approach using multi-mode microwave cavity resonance spectroscopy." Journal of Physics D: Applied Physics 52.3 (2018): 034004.). Due to the recent detection and investigation of an additional third decay regime of the afterglow of an extreme ultraviolet photon-induced plasma described in a later article (Platier, B., et al.\ "Transition from ambipolar to free diffusion in an EUV-induced argon plasma." Applied Physics Letters 116.10 (2020).) there is an additional reason for a minimum number of photons for this approach to work. Near or below this threshold, we explain that the response time of the diagnostic method is a limiting factor. Further, a second limit for the number of photons within a pulse is formalized related to the trapping of highly energetic free electrons.
\end{abstract}

\section*{Introduction}
\noindent In this addendum, two criteria for a sufficient number of photons within a radiation pulse for the beam monitor for ionizing radiation, introduced in an earlier article \cite{Beckers2019}, are explained. 
The underlying measurement principle of the monitor is Microwave Cavity Resonance Spectroscopy (MCRS). The first criterion is especially relevant for low-energy ionizing radiation and the later decay stages of plasmas induced by more energetic photons, while the second condition focusses on the trapping of highly-energetic free electrons. Both conditions are discussed below and come on top of the condition of sufficient resolution in the resonant behaviour.

\section*{First criterion: ambipolar diffusion}

In addition to the two decay regimes of the afterglow of an Extreme Ultraviolet (EUV) photon-induced plasma described in the literature \cite{Beckers2016b}, a third decay phase of this type of plasma has only recently been identified \cite{Platier2019Free}. This discovery was enabled by a dramatically improved lower detection limit of the governed diagnostics. In this third period, the decay rate of the electrons increases from the ambipolar rate up to the free diffusion rate. If an insufficient number of electrons are created by the photo-ionization events, triggered by the incoming photons, the transition to free diffusion immediately starts instead and the in-line monitor might not be able to react to the presence of the ephemeral electrons.

The electron density decays exponentially during the ambipolar phase and this rate increases during the transition towards free diffusion.
This transition is expected to set in when the following condition is satisfied \cite{Freiberg1968},
\begin{eqnarray}
\frac{\Lambda}{\lambda_\textrm{D}^\textrm{e}} = x \approx 100. \label{addendum:eq1}
\end{eqnarray}
Here, $\Lambda$ is the characteristic diffusion length of the chosen geometry and $\lambda_\textrm{D}^\textrm{e}$ is the electron Debye length,
\begin{eqnarray}
\lambda_\textrm{D}^\textrm{e}=\sqrt{\frac{\varepsilon_0 k_\textrm{B} T_\textrm{e}}{n_\textrm{e} e^2}}, \label{addendum:eq2}
\end{eqnarray}
where $\varepsilon_0$ is the vacuum permittivity, $k_\textrm{B}$ the Boltzman constant, $T_\textrm{e}$ the temperature of the electrons, $n_\textrm{e}$ the electron density and $e$ the elementary charge. For a finite cylinder $\Lambda$ \cite{Luikov1968} is,
\begin{eqnarray}
\Lambda=\left[\left(\frac{2.405}{r_\mathrm{cav}}\right)^2+\left(\frac{\pi}{h_\mathrm{cav}}\right)^2\right]^{-1/2},
\end{eqnarray}
where $r_\mathrm{cav}$ is the radius of the cylinder and $h_\mathrm{cav}$ the height of it.

An exact value of $x$ can be determined by studying the decay of the plasma and using the parameters of the exact moment of the start of the third decay regime. In the aforementioned study where the third decay regime was detected and investigated \cite{Platier2019Free}, the critical point $x$ was only as large as 6.

The simple linear relation for the maximum transient electron density $n_\textrm{e}^\textrm{max}$ used priorly \cite{Beckers2019,Ruud2015} is given by:
\begin{eqnarray}
n_\textrm{e}^\textrm{max} = n_\textrm{gas} \sigma N_\textrm{ph} m_\textrm{ion} m_\textrm{col}, \label{addendum:eq4}
\end{eqnarray}
where $n_\textrm{gas}$ is the number density of the neutral gas, $\sigma$ the photoionization cross-section at the specific photon energy, $N_\textrm{ph}$ the number of photons per pulse, $m_\textrm{ion}$ the multiplicator which takes in account for the process of single and double ionization and $m_\textrm{col}$ the multiplicator which compensates for further electron-impact ionization of neutrals by the electrons created by photonionization. Values for $m_\textrm{ion}$ and $m_\textrm{col}$ under lithography tool conditions are 1.05 \cite[p.~14]{Ruud2015} and at maximum 6 \cite[p.~15-16]{Ruud2015}, respectively.

By combining equations \ref{addendum:eq1}, \ref{addendum:eq2} and \ref{addendum:eq4}, the following condition for the minimum number of photons $N_\textrm{ph}^\textrm{min}$ is obtained:
\begin{eqnarray}
N_\textrm{ph}^\textrm{min} = \frac{\varepsilon_0 k_\textrm{B} T_\textrm{e} x^2}{n_\textrm{gas} \sigma m_\textrm{ion} m_\textrm{col} \Lambda^2 e^2}. \label{addendum:cond1}
\end{eqnarray}
Below this quantity, insufficient ionization takes place which may result in very short lifetimes of the electrons with respect to the fundamental response time of the cavity.

\section*{Second criterion: the potential well}

The first step in extending the lifetime of the electrons is the formation of a potential trap. This field can be formed when initial electrons reach the walls while the ions remain in the plasma. The approach of Van der Horst \cite[p.~71-73]{Ruud2015} will be used to estimate the net charge required for trapping the remaining free electrons. Note that in this approach, it is assumed that the collection of free electrons and ions---here referred to as the plasma---does not reach the cavity walls, the duration of a photon pulse is infinitely short and the mean free path of the electrons is much smaller than the typical dimensions of the plasma and the cavity.

An infinite long coaxial structure where the core with a radius $r_\textrm{p}$ is the volume occupied by the plasma and the outer region a vacuum region up to the cavity wall at a distance $r_\textrm{cav}$ 

To calculate the electric field $\mathbf{E}$, we apply Gauss's law:
\begin{equation}
\oint_A \mathbf{E} \cdot d\mathbf{a} = \frac{Q_\textrm{encl}}{\varepsilon_0},
\end{equation}
where $Q_\textrm{encl}$ is the charge enclosed by a surface $A$.

For the coaxial structure, the following equation for $\mathbf{E}$ (for $r$~$\geq$~$r_\textrm{p}$) can be obtained:

\begin{equation}
\mathbf{E} = \frac{\lambda_\textrm{q}}{2 \pi \varepsilon_0 r} \mathbf{e}_\textrm{r},
\end{equation}
where $r$ is the radial coordinate, $\mathbf{e}_\textrm{r}$ the unit vector in the radial direction, and $\lambda_\textrm{q}$ the charge density per unit length,
\begin{equation}
\lambda_\textrm{q} = e \Delta n_\textrm{i-e} \pi r_\textrm{p}^2.
\end{equation}
Here, $e$ is the elementary charge and $\Delta n_\textrm{i-e}$ the difference in the densities of the ions $n_\textrm{i}$, electrons $n_\textrm{e}$  $(\Delta n_\textrm{i-e}= Z n_\textrm{i}-n_\textrm{e})$, and $Z$ the average ionization stage of the ions.

Integration over $r$ results in a potential difference $\Delta V$ between the plasma and the walls:
\begin{equation}
\Delta V = - \int_{r_\textrm{cav}}^{r_\textrm{p}} \mathbf{E} \cdot d\mathbf{r} = \frac{e \Delta n_\textrm{i-e} r_\textrm{p}^2}{2 \varepsilon_0} \ln \left( \frac{r_\textrm{cav}}{r_\textrm{p}} \right).
\end{equation}

Once $e\Delta V$ is larger than the kinetic energy of the electrons $E_\textrm{ph-ion}$ directly after the photonionization process, the free electrons are trapped in the potential well. The required difference in charged species to satisfy this condition is given by:
\begin{equation}
\Delta n_\textrm{i-e} = \frac{2 \varepsilon_0 E_\textrm{ph-ion}}{e^2 r_\textrm{p}^2 \ln \left( \frac{r_\textrm{cav}}{r_\textrm{p}} \right)}.
\end{equation}

As the path of the electrons to the wall is much shorter than the mean free path, there is no need to correct for secondary ionization. Using a simplified version of equation \ref{addendum:eq4} for the maximum detected electron density, 
\begin{equation}
n_\textrm{e}^\textrm{max} = n_\textrm{gas} \sigma N_\textrm{ph} m_\textrm{ion}, 
\end{equation}
a second criterion for sufficient number of photons is formalized:
\begin{eqnarray}
N_\textrm{ph}^\textrm{min} \gg \frac{2 \varepsilon_0 E_\textrm{ph-ion}}{e^2 r_\textrm{p}^2 n_\textrm{gas} \sigma m_\textrm{ion} \ln \left( \frac{r_\textrm{cav}}{ r_\textrm{p}} \right)}. \label{addendum:cond2}
\end{eqnarray}
If this condition is met, the number of lossed electrons---which together with the remaining ions form the potential well---is neglectable in comparison to the maximum number of detected electrons.

During the oscillations in this well, the electrons lose energy due to collisions with the background gas. At low-pressure, the plasma will expand up to the walls of the cavity where the recombination takes place \cite[p.~77-78]{Ruud2015}.

\section*{Conclusion}

In summary, two conditions for the minimum number of photons (equations \ref{addendum:cond1} and \ref{addendum:cond2}) are introduced to prevent significant loss of free electrons before detection by the Microwave Cavity Resonance Spectroscopy based beam monitor for ionizing radiation \cite{Vadim2018}.

\section*{References}

\end{document}